# Detection of Topological Spin Textures via Non-Linear Magnetic Responses

*M. Stepanova[1,2,#], J. Masell[3,#], E. Lysne[1,2,#], P. Schoenherr[4,5], L. Köhler[6], Michael Paulsen[7], A. Qaiumzadeh[2], N. Kanazawa[8], A. Rosch[9], Y. Tokura[3,8,10], A. Brataas[2], M. Garst[6,11], and D. Meier[1,2,\**

[1] Department of Materials Science and Engineering, Norwegian University of Science and Technology (NTNU), 7491, Trondheim, Norway
[2] Center for Quantum Spintronics, NTNU, Trondheim, 7491, Norway
[3] RIKEN Center for Emergent Matter Science (CEMS), Wako, 351-0198, Japan
[4] School of Materials Science and Engineering, UNSW Sydney, Sydney, NSW 2052, Australia
[5] ARC Centre of Excellence in Future Low-Energy Electronics Technologies (FLEET), UNSW Sydney, Sydney, NSW 2052, Australia
[6] Institute of Theoretical Solid State Physics, Karlsruhe Institute of Technology, 76049 Karlsruhe, Germany
[7] Physikalisch-Technische Bundesanstalt (PTB), Berlin, 10587, Germany
[8] Department of Applied Physics, University of Tokyo, Tokyo, 113-8656, Japan
[9] Institute for Theoretical Physics, University of Cologne, Cologne, 50937, Germany
[10] Tokyo College, University of Tokyo, Tokyo, 113–8656, Japan
[11] Institute for Quantum Materials and Technology, Karlsruhe Institute of Technology, 76021 Karlsruhe, Germany

*corresponding author: dennis.meier@ntnu.no
#The authors contributed equally to this work.

**Topologically non-trivial spin textures, such as skyrmions and dislocations, display emergent electrodynamics and can be moved by spin currents over macroscopic distances. These unique properties and their nanoscale size make them excellent candidates for the development of next-generation logic gates, race-track memory, and artificial synapses for neuromorphic computing. A major challenge for these applications – and the investigation of nanoscale magnetic structures in general – is the realization of detection schemes that provide high resolution and sensitivity. We study the local magnetic properties of disclinations, dislocations, and domain walls in FeGe, and reveal a pronounced response that distinguishes the individual spin textures from the helimagnetic background. Combination of magnetic force microscopy and micromagnetic simulations links the non-linear response to the local magnetic susceptibility. Based on the findings, we propose a read-out scheme using superconducting micro-coils, representing an innovative approach for detecting topologically non-trivial spin textures and domain walls in device-relevant geometries.**

The discovery of magnetic skyrmions[1–3] and their emergent physical properties[4–8] propelled the research on topological spin states in solid state systems and motivated new concepts for spintronics devices where skyrmions act as mobile information carriers.[9-12] Skyrmions are intriguing as they are nanoscale objects that efficiently couple to spin currents, enabling high storage density and low-energy control.[3,5,13] With the progress of the field, the scope widened and other spin textures, such as merons,[14] biskyrmions[15] and hopfions[16,17] have been considered. Recently, disclinations, dislocations, and helimagnetic domain walls emerged as a new family of topological nano-systems that naturally arise in the helimagnetic ground state in chiral magnets.[18–20] The emergence of these topological spin textures is enabled by the lamellar-like morphology of the helimagnetic order analogous to, e.g., cholesteric liquid crystals,[21] swimming bacteria,[22] and the skin on our palms.[23] In magnetism, certain analogies exist to ferromagnetic stripe domains[24], but the involved length scales are substantially different. In chiral magnets, the spin structure twists continuously and the periodicity is up to three orders of magnitude smaller than for the classical stripe domains.[18] Edge dislocations within the helimagnetic structure are formed by a pair of $+\pi$ and $-\pi$ disclinations and – depending on their Burgers vector – can carry a topological charge $W = -\frac{1}{2}$. Such dislocations are topologically equivalent to half-skyrmions or merons as discussed in ref.[19]. Both disclinations and edge dislocations arise even without the external magnetic field usually needed to stabilize skyrmions and represent important building blocks for the formation of helimagnetic domain walls. It is now established that spin textures with non-trivial topology hold great technological potential enabling, e.g., reconfigurable logic gates,[11,25] race track memory,[9,10] and neuromorphic[26] / reservoir computing.[27] Sensing of individual topological spin structures, and magnetic nano-objects in general, in a way that is compatible with the proposed device architectures and semiconductor fabrication methods, however, remains a challenging task. Topological spin arrangements have been resolved by various imaging techniques, including electron,[9,28] X-ray,[29,30] and magneto-optical[29] microscopy, as well as scanning tunneling microscopy,[30,31] magnetic force microscopy (MFM),[32,33] and nitrogen vacancy magnetometry.[34] While these methods have provided important insight into the physics of topologically non-trivial spin textures, they are not directly transferrable to devices. For the specific case of skyrmion-electronics, a promising method is to utilize the topological Hall effect,[35,36] or magnetoresistance measurements,[37] but an expansion towards other magnetic nano-entities remains to be demonstrated. Thus, the development of dynamical and more agile read-out schemes that allow for resolving individual nanoscale spin textures in device-relevant geometries is highly desirable.

Here, we demonstrate how non-linear magnetic responses can be utilized to detect and identify both topologically trivial and non-trivial spin textures at the nanoscale. Combining MFM and micromagnetic simulations, we analyze the local magnetic response of magnetic disclinations, dislocations, and helimagnetic domain walls in the model system FeGe. Our results clarify the local magnetic properties and reveal characteristic fingerprints that enable selective detection of different nanoscale spin arrangements using superconducting micro-coils. The general feasibility and universality of this approach are demonstrated by two examples, considering the signal formation for an edge dislocation ($W = -\frac{1}{2}$) as well as a topologically trivial curvature domain wall ($W = 0$).

**Local magnetic response at dislocations and domain walls**

FeGe belongs to the family of helimagnets with B20 structure[2,3,38–40] and its magnetic phase diagram is well-established.[41] In addition, diverse nanoscale spin textures have been observed and their equilibrium structure has been analyzed in detail, rendering FeGe an ideal model system for this work. FeGe develops helimagnetic order below $T_N$ = 280 K, stabilized by the competition between Heisenberg exchange and the relativistic Dzyaloshinskii-Moriya interactions in the non-centrosymmetric crystal lattice. The helimagnetic ground state is characterized by a gradual rotation of the magnetization vector $M$ about a wave vector $q = (\frac{2\pi}{\lambda})\hat{q}$, where $\lambda$ = 70 nm[42] is the helical period and $\hat{q}$ characterizes the direction of the helical axis (Fig. 1a).

Fig. 1b shows an MFM image of FeGe in the helimagnetic phase, recorded in two-pass mode (see Experimental Section for details). Bright and dark lines reflect the periodic magnetic structure with $\lambda \approx 70$ nm and $q$ lying in the surface plane (white arrow) in agreement with neutron scattering data.[42] Because of the lamellar morphology[18,19] – which is analogous to cholesteric liquid crystals – different types of defects arise in FeGe at the nanoscale, including topologically non-trivial objects such as dislocations (Fig. 1c) and zigzag disclination walls (Fig. 1e), respectively, as well as more simple curvature walls that do not carry a topological charge (Fig. 1d). Details about the inner structure of the different spin textures are reported elsewhere.[18,19] Most importantly for this study, Fig. 1c-e reveal a universal feature that is shared by all defect structures, independent of their topology, shape and dimensionality: All defects exhibit an additionally enhanced contrast in the MFM data that is not observed in regions with perfect lamellar-like order (Fig. 1b), separating them from the helimagnetic background. A similar MFM response has been observed at the helimagnetic domain walls in earlier studies, but without clarifying the microscopic origin.[19] Thorough examination of the data in Fig. 1 reveals that the enhanced MFM response is asymmetric: only the bright lines, which indicate an attractive force between probe tip and sample, exhibit increased intensity and width. Furthermore, we find that the enhanced MFM signal can be detected more than 100 nm above the surface, that is, before the actual spin structure of the defects is resolved (see Supplementary Fig. **1** and Supplementary Fig. **2**).

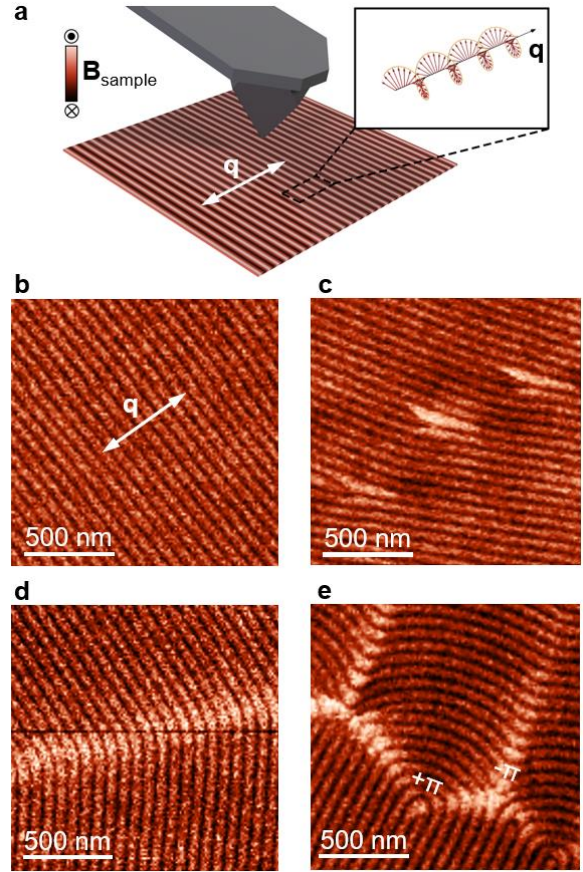

**Figure 1. MFM imaging of helimagnetic order, dislocations and domain walls. a**, Schematic illustration of the helical spin structure in FeGe described by the wave vector $q$ and the characteristic stripe-like pattern probed by MFM in the helical phase. The colour scale indicates the direction of the magnetic stray field $B$ from the sample. **b**, MFM image of the helimagnetic order within a single $q$ domain in FeGe. Note that the bright and dark contrast originate from the spin helix, giving rise to a lamellar morphology with a measured periodicity of ≈ 70 nm, which is about three orders of magnitude smaller than the conventional stripe domains in ferromagnetic systems. In FeGe, domains are formed only on much larger length scales[19] as seen, e.g., in (**d**) and (**e**), corresponding to regions with a different orientation of $q$. **c – e**, MFM images showing magnetic dislocations in the lamellar-like spin structure (**c**), a curvature domain wall (**d**), and a zigzag domain wall composed of +π and −π disclinations (**e**). All 1D and 2D spin textures in (**c**) to (**e**) exhibit enhanced bright MFM contrast compared to the helimagnetic background.

**Micromagnetic simulations**

In order to understand the unusual local response of the magnetic defects, we conduct micromagnetic simulations. The MFM signal is proportional to the phase shift $\Delta\phi \propto -\partial^2 E_{int}/\partial z_0^2$ of the oscillating probe tip. $E_{int}$ is the dipolar interaction energy between the tip and sample and $z_0$ denotes the tip-sample distance. First, we theoretically discuss $E_{int}$ on the level of linear response where the tip just acts as a probe and its influence on the magnetization of the sample is neglected:

$$E_{int} = -\int d\mathbf{r} d\mathbf{r}' \mathbf{M}_{tip}(\mathbf{r}) \chi_d^{-1}(\mathbf{r} - \mathbf{r}') \mathbf{M}(\mathbf{r}') \quad (1)$$

Here, $\mathbf{M}_{tip}$ is the magnetization of the MFM tip, $\mathbf{M}$ is the magnetization in the sample, and $\chi_d^{-1}$ is the dipolar interaction between the tip and the sample (see Supporting Note S1). $\mathbf{M}_{tip}$ is often approximated by a point-dipole ($\mathbf{M}_{tip}(\mathbf{r}) = \mathbf{m}_{tip}\delta^3(\mathbf{r} - \mathbf{r}_0)$; $\mathbf{m}_{tip}$ is the magnetic moment of the tip and $\mathbf{r}_0$ its position). The stray field $\mathbf{B}$ of the helical phase is shown in Fig. 2a. It decreases exponentially as function of the distance[18], $\mathbf{B} \propto e^{-2\pi z_0/\lambda}$, with $\lambda/2\pi$ determining the decay length of the stray field, see Supplementary Note. Microscopically, $\mathbf{B}$ is generated by magnetic bulk charges, $\nabla \cdot \mathbf{M}$, and surface charges, $\mathbf{n} \cdot \mathbf{M}$ ($\mathbf{n}$ is the unit vector normal to the surface). As bulk charges are absent in the case of an ideal spin helix, $\mathbf{B}$ is dominated by surface charges, as shown in Fig. 2b. Thus, taking into account that the projected periodicity $\lambda_p$ of the lamella-like order increases whenever the helimagnetic structure is bent, i.e. $\lambda_p > \lambda$, the generally stronger magnetic stray field $\mathbf{B}$ and phase shift $\Delta\phi$ at defects can be explained based on an increased decay length ($\sim e^{-2\pi z_0/\lambda_p}$) associated with local magnetic surface charges (see Supplementary Fig. 3, Supplementary Fig. 4, Supplementary Note). However, while this geometrically driven effect can give rise to higher attractive and repulsive forces at defect structures, it is qualitatively different from the asymmetric effect presented in Fig. 1, failing to explain why only attractive forces appear amplified at the defect sites.

This discrepancy leads us to the conclusion that the dipolar interaction between the magnetic moment of the tip and the stray field of the spin helix is non-negligible.[24] Therefore, we go beyond the linear response theory that describes non-invasive MFM measurements and include the local stray field of the magnetic MFM tip in our three-dimensional simulations, accounting also for emergent non-linear responses. We model the tip by a single dipole moment and the magnetization of FeGe is described by a lowest order gradient expansion:

$$E[m] = \int_V \left[A(\nabla \mathbf{m})^2 + D\,\mathbf{m}\cdot(\nabla \times \mathbf{m}) - \mathbf{M}\cdot\mathbf{B}_{tip}\right] dV \quad (2)$$

subject to the boundary conditions of the embedding helical phase in three spatial dimensions ($\mathbf{m} = \mathbf{M}/M_s$ is the normalized magnetization, $M_s$ is the saturation magnetization, $A$ is the exchange stiffness, $D$ is the Dzyaloshinskii-Moriya interaction, and $\mathbf{B}_{tip}$ is the dipolar stray field that is emitted by the tip). For simplicity, we neglect effects from the demagnetizing field and cubic anisotropies as they do not change the results qualitatively (see Methods for details on parameters and software). The calculated helimagnetic texture resulting under realistic experimental conditions is presented in Fig. 2c. As function of distance $z_0$, the stray field from the tip decays as $\mathbf{B}_{tip} \propto z_0^{-3}$ in the dipole approximation; see Supplementary Note for a discussion of pyramidally shaped tips. As a consequence, the strongest polarizing effects in the spin helix are observed at the sample surface, where the tip induces an additional magnetic surface charge (blue in Fig. 2d). Vice versa, we find that the net induced surface charge is the main source for the magnetic stray field probed by the tip (bent black lines in Fig. 2c), leading to a stronger and more long-ranged attractive force (polynomially decaying instead of exponentially) compared to the unperturbed helimagnetic structure as displayed in Fig. 2a. Note that this net polarization of the magnetization occurs for helimagnetic order both with and without defect structures (see Supporting Note S1 for details). In summary, the magnetization of the tip $\mathbf{M}_{tip}$ leads to an additional non-linear response in MFM. In a second order process, $\mathbf{M}_{tip}$ creates magnetic charges within the sample via dipolar interactions which then feed back onto the tip:

$$E_{int}^{(2)} = -\frac{1}{2}\int d\mathbf{r} d\mathbf{r}' d\mathbf{r}_1 d\mathbf{r}_2 \mathbf{M}_{tip}(\mathbf{r}) \chi_d^{-1}(\mathbf{r} - \mathbf{r}_1)\chi(\mathbf{r}_1, \mathbf{r}_2)\chi_d^{-1}(\mathbf{r}_2 - \mathbf{r}') \mathbf{M}_{tip}(\mathbf{r}') \quad (3)$$

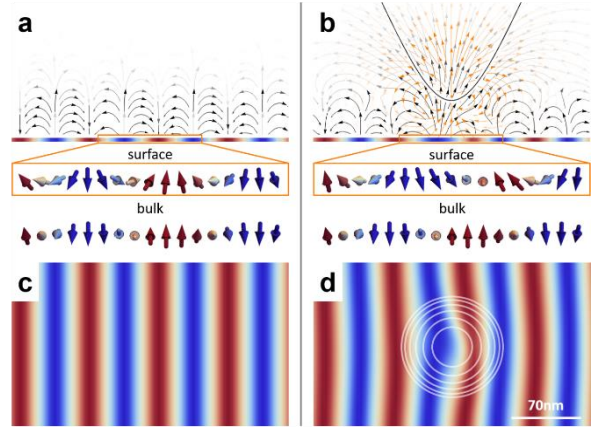

**Figure 2. Calculated local response of the helimagnetic spin structure.** **a**, Side view of the helimagnetic order which – in the absence of an invasive magnetic tip – differs only slightly for the surface (top) and the bulk (bottom). The out-of-plane magnetic components associated with the helical spin structure (sketched by solid black arrows) generate alternating magnetic surface charges (blue to red) with a periodicity of about 70 nm, which are the main source for the magnetic stray field (curved black arrows, saturation encodes field strength). **b**, same as in (**a**) in the presence of an invasive magnetic tip ($\mathbf{m}_{tip} = 10^{-16}$ A m²), positioned at distance $z_0 = 40$ nm from the surface), leading to substantial changes in the helimagnetic structure at the surface (top) compared to the bulk (bottom). The extra field of the tip in (**b**), approximated by a single dipole, is coloured orange. **c**, top view of the alternating magnetic surface charges seen in (**a**). **d**, same as in (**c**) in the presence of an invasive magnetic tip. The position of the MFM tip is indicated in (**d**) by white rings, each corresponding to a factor 2 decreased magnetic field. The polarizing influence of the tip in (**b**), (**d**) is clearly visible.

The efficiency for inducing magnetic charges, however, is set by the magnetic susceptibility $\chi$, so that local variations in $\chi$ can lead to additional contributions in MFM. At the level of domains, such additional contributions have been studied intensively and are known as susceptibility contrast.[24] Our calculations reveal that such susceptibility contributions are equally important at the nanoscale, leading to substantially different non-linear responses for spin-helix segments with different out-of-plane magnetization components.

**Non-linear response at helimagnetic domain walls**

To verify that the local susceptibility contrast observed at helimagnetic defects originates from tip-induced magnetic surface charges – i.e., a non-linear response – we simulate MFM scans with oppositely magnetized tips. As an instructive example, we consider the case of a topologically non-trivial zigzag wall containing +π and -π disclinations.[19] For reference, the magnetic surface charges of the ideal, undisturbed zigzag domain wall are presented in Fig. 3a. Simulations accounting for the tip-sample interaction are presented in Fig. 3b,d. The simulations show that the enhanced magnetization of the +π and -π disclination centres inverts as the magnetization of the tip is switched from "down" (Fig. 3b) to "up" (Fig. 3d). In contrast, the lines connecting the +π and -π disclinations remain bright upon reversal, corresponding to an attractive force on the tip independent of the orientation of $M_{tip}(r)$. Corresponding experimental MFM scans on a real sample with "down" and "up" magnetized tip are presented in Fig. 3c and e, respectively (see Methods for experimental details). Both MFM images in Fig. 3c reveal qualitatively the same spin texture at the zigzag wall. In agreement with the simulations, we observe that the MFM signal at the disclination centre inverts (marked by white dashed circles). The signal associated with the domain wall, however, remains bright in both scans (marked by white dashed lines to the right of the circles). The data thus confirms an attractive force that occurs at the site of the domain wall due to tip-induced magnetic surface charges in the helical spin structure, leading to pronounced susceptibility contrast in MFM.[24,43-45] As the magnetization at defects deviates from the energetically favourable helical structure, it is reasonable to assume that it is more susceptible to external magnetic fields. The higher susceptibility leads to a more efficient generation of surface charges than in the helimagnetic background and, hence, an additional long-ranged attractive force, consistent with our MFM measurements (Supplementary Fig. 1, 2, 5-8). In contrast to the MFM contributions from magnetic surface charges, however, the local susceptibility contrast is proportional to $m_{tip}^2$ (Supplementary Note). As a consequence, susceptibility-related signals linked to the zigzag wall do not invert along with the tip magnetization and can be isolated by adding MFM images gained with opposite tip magnetization. The latter is presented in Supplementary Fig. 9, 10 and 11 where the MFM sum image reveals a pronounced contrast associated with the curved helix structure, confirming that defects in the helimagnetic structure in FeGe exhibit a locally enhanced susceptibility.

**Generalization of the detection scheme**

In summary, we have identified a pronounced non-linear response at defects of helimagnetic order in FeGe due to their specific surface polarization induced by the magnetic tip (Fig. 2). This non-linear response occurs as a magnetic field is applied, providing an additional opportunity for sensing and distinguishing nanoscale spin textures. In Figure 4, the latter is shown for selected examples of 1D- and 2D-defects; that is, a magnetic edge dislocation (topologically equivalent to a half-skyrmion or meron) and a topologically trivial curvature wall. The basic detection scheme is presented in Fig. 4a, displaying an artistic view of a SQUID (superconducting quantum interference device) coil with a diameter of r = 500 nm comparable to state-of-the-art SQUID-on-tip technology.[46,47] Such nanoSQUIDs can be operated both without and with magnetic background fields, facilitating a sensitivity of 0.6 $\mu_B$ Hz$^{-1/2}$ at 1 T as demonstrated in ref.[48] Thus, using SQUIDs it is feasible to measure both linear and non-linear responses, detectable as a variation in the magnetic flux ΔΦ. To quantify and compare the expected signals, we assume an idealized geometry where a magnetic field of 100 mT is produced by the same coil as used for detection (Fig. 4b-e). We note, however, that this field strength may be technologically challenging to realize with such SQUIDs. A conceptually equivalent alternative is to generate the local background field via patterned magnetic elements.[49] The calculations show that for a magnetic dislocation, which binds a finite magnetic surface charge, both linear (0 mT) and non-linear (100 mT) detection is possible, yielding comparable changes in the magnetic flux (Fig. 4c). The 180° phase jump, which is induced as dislocations move through the helimagnetic background,[18] however, is more pronounced when applying non-linear detection. In contrast to edge dislocations, curvature domain walls exhibit alternating surface charges that macroscopically average to zero, so that the linear signal can effectively cancel out. However, these defects are then still detectable via the non-linear interaction (Fig. 4e) which generates a clear peak at the domain wall. This peak makes the signal asymmetric so that it remains detectable even when its oscillating fine-structure cannot be resolved. Based on the calculations, a signal span in magnetic flux of $\Delta\Phi \sim 10^{-16}$ Wb $\approx 0.2\,\Phi_0$ is expected. For the coil in Fig. 4, this difference translates into a magnetic field $\Delta\Phi / \pi r^2 \approx 510$ µT, which is readily measurable using SQUID magnetometers. Importantly, the variation in magnetic flux presented in Fig. 4c,e is specific to the cases depicted in Fig. 4b,d; in general, the measured signal depends on the coil geometry and position, as well as the direction of movement relative to the spin texture, providing additional information about the magnetic order at the nanoscale. For example, the magnetic flux is constant for translations perpendicular to the **q**-vector of the spin helix, but varies in the direction parallel to **q**, allowing to resolve phase jumps (Fig. 4c) and spatial modulations (Fig. 4e) in the helimagnetic spin structure.

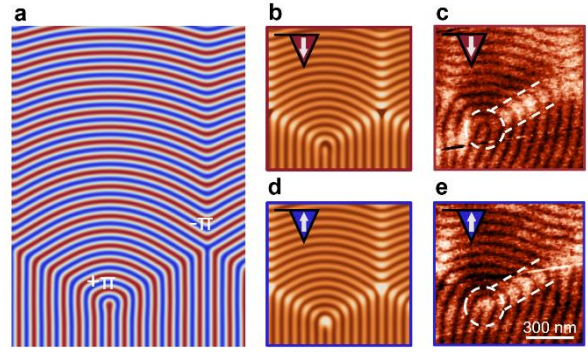

**Figure 3. Magnetic response from a zigzag domain wall with alternating +π and −π disclinations as function of the orientation of the tip magnetization. a**, Calculated magnetic surface charges of a zigzag domain wall at the surface of FeGe. The colour denotes the out-of-plane magnetization related to the spin helix from pointing up (blue) to down (red). **b, d**, Calculated non-linear MFM response for a down (red, b) and up (blue, d) magnetized tip, taking the tip-sample interaction into account (lift height: 100 nm, tip moment: 2 · $10^{-16}$ A m²). Bright and dark colours indicate attractive and repulse forces, respectively. The pattern of bright and dark lines associated with the spin helix inverts as the tip changes magnetization direction, whereas an additional attractive force is detected at the domain wall position independent of the tip magnetization. **c, e**, Corresponding MFM images of a zigzag domain wall recorded at the same position with (c) tip magnetized down and (e) up. The size of the scanning area is 1 μm × 1 μm. The white dashed circles mark the centre of the disclination, and the white dashed lines mark the domain wall.

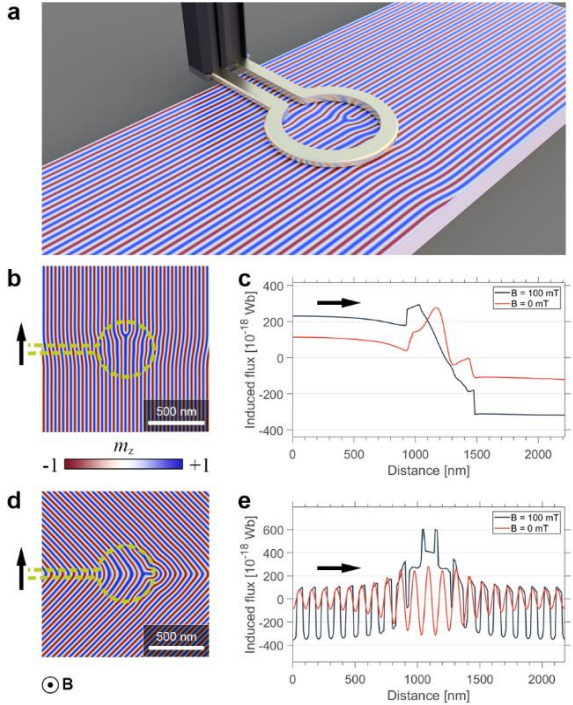

**Figure 4. SQUID-based read-out scheme for the detection of 1D and 2D magnetic spin textures. a**, Illustration of a magnetic track and schematic SQUID coil (diameter: 500 nm) positioned 10 nm above the surface, presenting the basic setup for detection. Using stationary coils, mobile topological spin textures – here, a dislocation – can be sensed and counted via a defect-specific change of the magnetic flux through the coil. **b**, Out-of-plane magnetization, $m_z$, of an edge dislocation under the influence of the magnetic stray field (100 mT) from the coil illustrated by the yellow dashed line. **B** gives the direction of the magnetic field within the coil and the black arrow indicates the direction of motion relative to the edge dislocation. **c**, Induced magnetic flux measured with biased (100 mT) and non-biased (0 mT) coils. **d**, and **e**, Same as in (b) and (c) for a curvature wall.

On the one hand, the proposed detection scheme is compatible with local imaging techniques such as scanning SQUID microscopy,[50] where the coil is scanned across the sample surface to detect the defects, removing the requirement of sub-100 nm resolution to verify emergent defect structures. On the other hand, race-track like geometries[9,51] with stationary coils are possible, sensing moving edge dislocations, domain walls and other mobile topological defects via their non-linear response.

## Conclusions

The results presented in this work clarify the interaction of topologically non-trivial spin textures and domain walls in chiral magnets with external magnetic field, revealing a pronounced non-linear response due to field induced surfaces charges. The additional surface charges lead to a more long-ranged interaction compared to the unperturbed magnetic state, facilitating new opportunities for the detection and differentiation of nanoscale spin textures using nanoSQUIDs. Based on the non-linear response, the detection sensitivity can be improved as demonstrated for two instructive examples, i.e., a magnetic edge dislocation and a helimagnetic curvature wall. The proposed detection scheme is universal and, in principle, can be applied to all magnetic nano-objects that exhibit a different susceptibility than their surroundings, enabling sensing of otherwise hidden 1D and 2D spin textures. The latter is supported by the recent observation of enhanced magnetic susceptibility at antiferromagnetic domain walls in the topological insulator MnBi$_2$Te$_4$,[45] expanding application opportunities into the realm of antiferromagnetic spintronics. Thus, in addition to the fundamental insight into the nanoscale physics of chiral magnets, this work introduces a viable read-out scheme for topological magnetic defects, and local spins arrangements in general, opening new possibilities for the design and fabrication of on-chip devices for spintronics.

## Methods

**Sample Preparation:** Single crystals of FeGe were grown by the chemical vapor transport method. To achieve flat high-quality surfaces for MFM imaging, the samples were prepared by lapping and polishing to achieve a root mean square roughness of approximately 1 nm and cleaned with high-purity acetone and methanol, following the same procedure as described in Refs.[18,19].

**Magnetic force microscopy:** The MFM data was recorded using a commercial SPM system (NT-MDT NTEGRA Prima AFM). Magnetic probe tips (PPP-MFMR from Nanosensors) with force constant of 2.8 N/m and quality factor $Q$ of about 200 were used. The tips possess a hard magnetic coating with effective magnetic moment of $10^{-16}$ A m$^2$ and have been magnetised by a permanent magnet prior to the measurements. Sample cooling was achieved using a water-cooled Peltier element and the measurements were carried out in N$_2$ atmosphere (with a few mBar overpressure) to prevent ice formation. The microscope was operated in two-pass MFM mode with the magnetic tip oscillating at its resonance frequency ($\approx$ 70 kHz) with an amplitude of $\approx$ 30 nm. During the first pass, a topography image was collected by scanning with the tip close to the sample surface. During the second pass, the tip was lifted 10-200 nm (in addition to 30 nm in the first pass) and retraced the measured topography to sense solely the magnetic interaction between the magnetic stray field of the sample and the magnetic tip. Between measurements with opposite magnetisation of the tip, the sample was heated to room temperature and the protective hood was removed to switch the tip magnetisation using a permanent magnet. Regions of interest were tracked using an optical microscope and the obtained MFM images were aligned using topographical features on the sample surface.

**Simulations:** We model the magnetization in FeGe in the presence of an MFM tip in an effective isotropic model. We assume that the magnetization of the MFM tip and hence also the stray field are constant. This model is similar to the micromagnetic model, but neglects contributions of the demagnetizing field within the sample as these significantly slow down our calculations while their corrections are assumed to be only marginal. The simulations were performed at $T = 0$ K using the following parameters for FeGe: $A = 8.78$ pJ m$^{-1}$, $D = 1.58$ mJ m$^{-2}$, and $M_s = 384$ kA m$^{-1}$.[52] For the numerics, we discretize the continuum theory on a regular mesh of cuboids $(a_x, a_y, a_z)$ with $a_i \approx \lambda/16$ where $\lambda = 4\pi A/D$ is the wavelength of the helix. The rather coarse discretization is still suitable as we approximate derivatives by fourth order stencils.[53] The energy of the 3d setup is minimized by a (single precision) GPU-accelerated self-written software[20] where one boundary condition is von-Neumann (the surface to vacuum) and the opposite boundary is fixed to the bulk minimizer. The other boundaries are also fixed but were pre-relaxed under the constraints of one "bulk" and one "surface" edge. The thickness of the 3d slab is usually of the order $L_z \approx 2\lambda$ with additional tests for $L_z \approx 8\lambda$. For one pixel in a non-linear MFM-image, we relax the magnetization in a local spheroid for three different tip heights which we then use to compute the second derivative of the dipolar interaction energy.

## Acknowledgements
M.S., E.L. and D.M. acknowledge funding from the Research Council of Norway, project number 263228, and support through the Norwegian Micro- and Nano-Fabrication Facility, NorFab (project number 295864). M.S., E.L., D.M., A.Q. and A.B. acknowledge support by the Research Council of Norway through its Centres of Excellence funding scheme, Project No. 262633, "QuSpin". D.M. thanks NTNU for support via the Onsager Fellowship Program and the Outstanding Academic Fellows Program. J.M. was financially supported as a Humboldt/JSPS International Research Fellow (19F19815). N.K. acknowledges funding from JSPS KAKENHI (Grant number JP20H05155). A.R. acknowledges financial support by the Deutsche Forschungsgemeinschaft (DFG) within CRC 1238 (project number 277146847, subproject C04). N.K. and Y.T. acknowledge funding from Core Research for Evolutional Science and Technology (CREST), Japan Science and Technology Agency (JST) (Grant No. JPMJCR1874). M.G. is supported by DFG SFB1143 (Project-id. 247310070), DFG Project-id. 270344603 and 324327023.

## Author contributions
M.S., E.L. and J.M contributed equally to this work. M.S., E.L., and P.S. conducted the MFM measurements under the supervision of D.M. L.K, J.M. and M.G. proposed the theoretical explanation. J.M. performed the micromagnetic simulations. A.R., A.Q., and A.B. analytically calculated the signals. N.K. and Y.T. provided the samples for this study. M.S., E.L., J.M., and D.M. developed the idea for the proposed read-out scheme in collaboration with M. P. M.S., J.M., and D.M. wrote the manuscript. All authors discussed the results and contributed to the final version of the manuscript.

**Supporting Information**

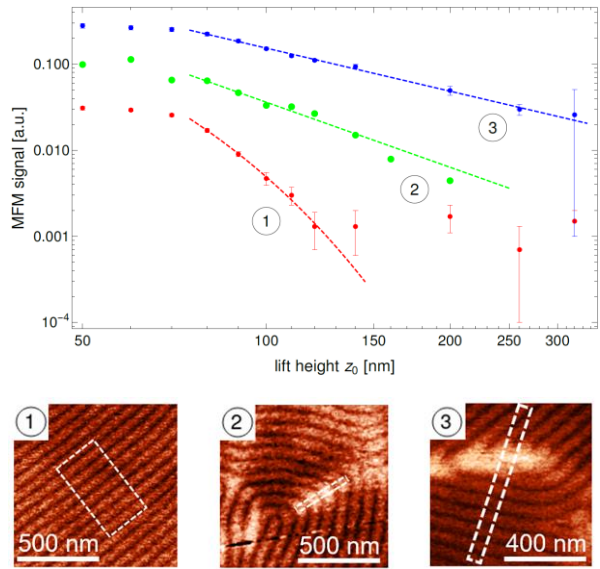

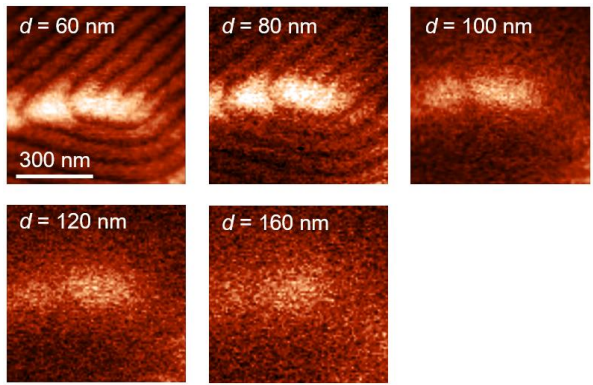

**Supplementary Fig. 1. Relation between MFM signal and lift height of the MFM tip.** Red data points represent the evolution of the helix amplitude, evaluated based on an MFM images series recorded with varying lift height of the probe tip. Data points are derived by fitting a sinusoidal function to averaged data taken from the region marked by the dashed box in (1). The red dashed line is a fit with an exponential decay $I = I_0\, e^{-\frac{2\pi}{L} z_0}$ with $L = 100 \pm 10$ nm. Green data points show the evolution of the response at the zigzag domain wall (type II) in (2). Plotted is the peak value of the MFM phase signal, calculated as the difference between an averaged profile from the dashed box in (2) and the averaged signal from the adjacent helimagnetic background. The green dashed line is a polynomial fit $I \propto z_0^{-\alpha}$ with $a = 2.5 \pm 0.5$. The blue data points are the peak value measured at the type III domain wall in (3), estimated by fitting a Gaussian function to the profile averaged over the dashed box in (3). The blue dashed line is a polynomial fit $I \propto z_0^{-\alpha}$ with $a = 1.66 \pm 0.33$.

**Supplementary Fig. 2. MFM scans recorded at a type III domain wall with increasing lift height.** For smaller tip-sample distances (60 nm and 80 nm) both magnetization and susceptibility contrast are resolved, whereas for larger distances ($\gtrsim 100$ nm) only the susceptibility contrast is detected.

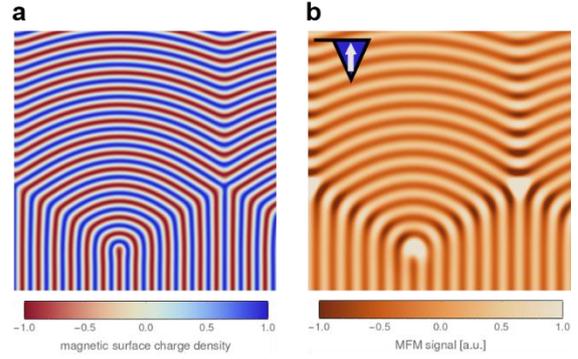

**Supplementary Fig. 3. Linear MFM signal in simulations. a**, Distribution of magnetic surface charges, i.e., out-of-plane components of the magnetization, of a zigzag domain wall at the surface of a bulk crystal. **b**, Simulated linear MFM signal (lift height: 100 nm, tip moment: $2 \cdot 10^{-16}$ A m$^2$, same as in Fig. 3b in the main text). The signal is enhanced at bent helices, both in the positive and negative direction. The color code is chosen such that the signal of the unperturbed helix oscillates between $\pm \frac{2}{3}$.

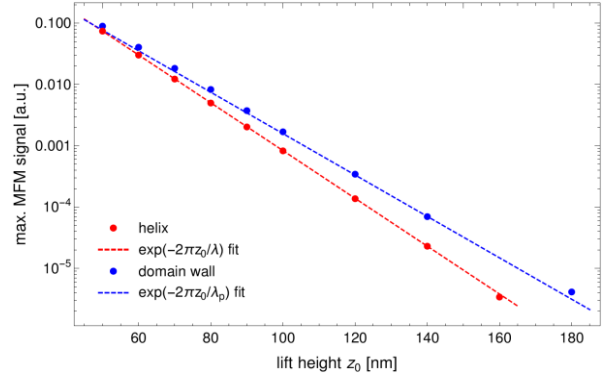

**Supplementary Fig. 4. Distance-dependence of simulated linear MFM signals.** Red dots show the maximal signal on a helix without defects. Blue dots show the maximal signal on a curvature domain wall with an angle $\varphi = 30°$ between the wall and the helical **q**-vector on either side. The tip-sample interaction is neglected. The data is well described by an exponential decay $I \propto e^{-\frac{2\pi}{\lambda} z_0}$ where $z_0$ is the lift height and $\lambda$ is the local wavelength, i.e., $\lambda = 70$ nm for a helix without defects and $\lambda = \lambda_p$ 70 nm / $\cos(\varphi)$ for the curvature wall, see dashed lines.

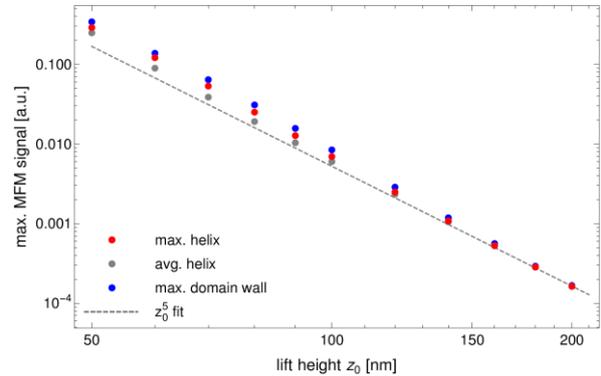

**Supplementary Fig. 5. Distance-dependence of simulated non-linear MFM signals.** Red dots show the maximal signal on a helix without defects. Gray dots show the average signal on the same helix. Blue dots show the maximal signal on a curvature domain wall with an angle $\varphi = 30°$ between the wall and the helical **q**-vector on either side. The tip is modelled as a dipole with moment $10^{-16}$ A m$^2$. In contrast to Supplementary Fig. 4, the average signals do not vanish. Instead, the average signal on the helix is well described by a power law $I \propto z_0^5$, see Supplementary Note for details. Deviations from this trend are discussed in Supplementary Fig. 6.

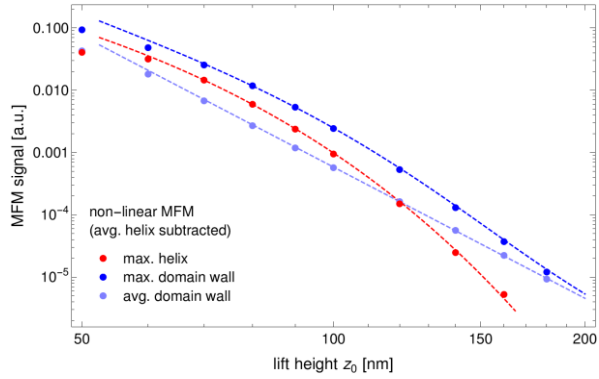

**Supplementary Fig. 6. Distance-dependence of simulated non-linear MFM signals.** The average signal in the helical phase, see Supplementary Fig. 5, has been subtracted. Red dots show the maximal signal on a helix without defects. Blue dots show the maximal signal on a curvature domain wall with an angle $\varphi = 30°$ between the wall and the helical $q$-vector on either side. Light blue dots show the average signal on this domain wall. The tip is modelled as a dipole with moment $10^{-16}$ A m². The signal on the helix is well described by an exponential fit $I = I_0\, e^{-\frac{2\pi}{\lambda}z_0}$ where $\lambda = 70$ nm is the wavelength of the helix, see red dashed line, in agreement with the discussion of the magnetic signal in the Supplementary Note, see supplementary equation (7). The average signal on the domain wall is well fitted by a power law $I \propto z_0^7$ which describes the susceptibility contrast. The discrepancy to the analytical prediction in the Supplementary Note, see supplementary equation (11), probably arises because we simulate a slab of finite thickness and not an infinitely thick specimen. Finally, the maximal signals on the domain wall are well fitted by the power law for the susceptibility contrast plus an additional exponential decay $\propto e^{-\frac{2\pi}{\lambda/\cos(\phi)}z_0}$ for the magnetic signal, see Supplementary Note for details.

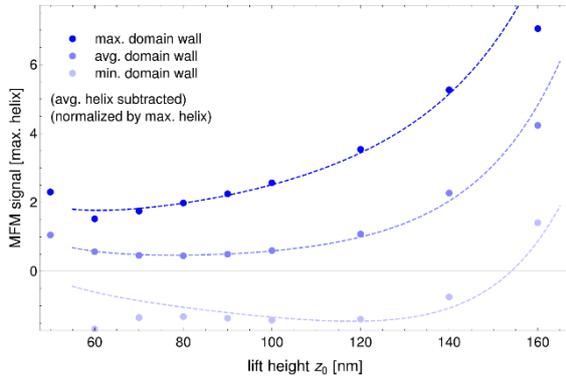

**Supplementary Fig. 7. Distance-dependence of simulated non-linear MFM signals.** The average signal in the helical phase, see Supplementary Fig. 5, has been subtracted. Data points show the maxima/average/minima of the MFM signal on a curvature domain wall ($\varphi = 30°$) normalized by the maximal signal on the helical phase without defects. The tip is modelled as a dipole with moment $10^{-16}$ A m². The dashed lines are the same fits as in Supplementary Fig. 6 (for the minima we just inverted the sign of the extra exponential decay on the domain wall). The graph illustrates how the signal on the domain wall becomes increasingly brighter than the signal on the helix as the distance between the tip and the sample is increased.

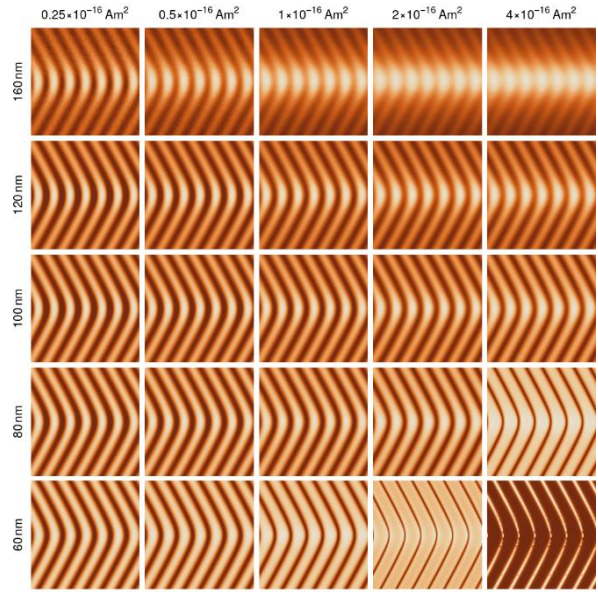

**Supplementary Fig. 8. Tip-dependence of simulated non-linear MFM signals.** The panel matrix shows MFM maps for various lift heights (rows) and tip magnetizations (columns), simulated for a curvature domain wall with an angle $\varphi = 30°$ between the wall and the helical $q$-vector on either side. Lift heights and tip moments are indicated in the figure.

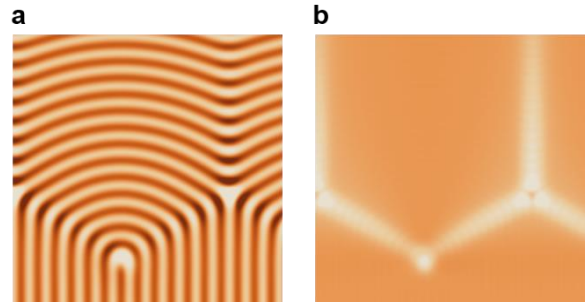

**Supplementary Fig. 9. Separation of the non-linear MFM signal in simulations into magnetization and susceptibility contrast. a**, magnetization contrast and **b**, susceptibility contrast, based on Fig. 3b in the main text. The magnetization contrast is defined as $I_m = (I_{up} - I_{down})/2$ and the susceptibility contrast as $I_\chi = (I_{up} + I_{down})/2$, where $I_{down}$ and $I_{up}$ are the contrasts for a "down"-polarized and "up"-polarized tip, c.f. Fig. 3b and d in the main text, respectively. See Supplementary Note for details.

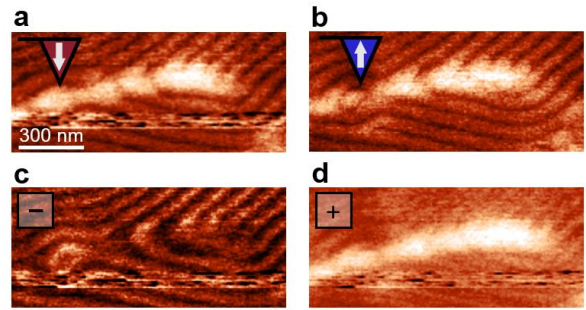

**Supplementary Fig. 10. Experimental separation of magnetization and susceptibility contrast. a**, and **b**, MFM images of the same region measured with the tip magnetized "down" and "up", respectively. The area shown here corresponds to the region marked by the white dashed line in the larger MFM scan presented in Supplementary Fig. 11. **c**, Magnetization contrast image gained by taking the difference of the data presented in (**a**) and (**b**), corresponding to $I_m$. **d**, Sum of the MFM data in (**a**) and (**b**), showing the susceptibility contrast image ($I_\chi$).

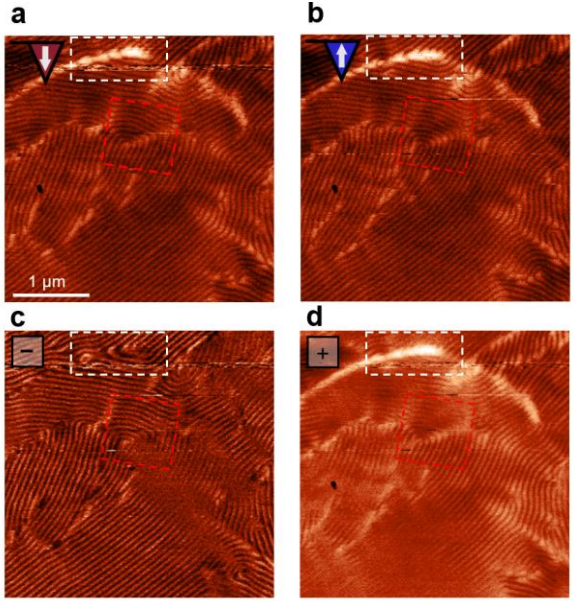

**Supplementary Fig. 11. Overview MFM images taken with oppositely magnetized tip. a**, and **b**, MFM images of the same region (3.8 μm × 3.8 μm) measured with the tip magnetized "down" and "up", respectively. To change the magnetization of the tip, the sample was transiently heated to room temperature, i.e., $T > T_N$. The region discussed in Fig. 3 in the main text is marked by the red dashed line and the area presented in Supplementary Fig. 10 is marked by the white dashed line. **c**, Magnetization contrast image gained by taking the difference of the data presented in **a** and **b**, corresponding to $I_m$. **d**, Sum of the MFM data in **a** and **b**, showing the susceptibility contrast $I_\chi$. In specific regions where the local spin texture is restored almost completely after heating above $T_N$ (most likely due to structural defects that act as pinning sites), a clear separation into magnetization and susceptibility contributions is possible (see white dashed box in **c** and **d**). In general, however, the spin textures observed before and after $T_N$ are not identical, leading to blurring as MFM images are added or subtracted. Despite this effect, the experiments reveal that helix-related MFM contrasts invert along with the tip magnetization as explained in detail in the main text, considering the region marked by the red dashed box.

## Supplementary Note S1. Susceptibility contrast and magnetization contrast in MFM

**Formulation of the non-linear problem.** When computing the energy of a magnetic tip, one has to consider that the presence of the tip leads to a distortion of the magnetization of the sample. If this distortion is small, it can be calculated from the susceptibility $\chi(\mathbf{r}, \mathbf{r}')$ of the sample and the dipolar field $\int d\mathbf{r}' \chi_d^{-1}(\mathbf{r} - \mathbf{r}')\mathbf{M}_{tip}(\mathbf{r}')$ due to the magnetization $\mathbf{M}_{tip}(\mathbf{r}')$ of the tip with

$$\left(\chi_d^{-1}(\mathbf{r})\right)_{ij} = \frac{\mu_0}{4\pi}\frac{3 r_i r_j - \delta_{ij} r^2}{r^5} = \frac{\mu_0}{4\pi}\partial_{r_i}\partial_{r_j}\frac{1}{r}. \quad (S1)$$

The total change of magnetic energy due to the presence of the tip can then be approximated as

$$U \approx U_m + U_\chi \quad (S2)$$

where the two contributions are sensitive to the magnetization $\mathbf{M}$ and susceptibility $\chi$ of the sample, respectively, with

$$U_m = -\int d\mathbf{r} d\mathbf{r}' \mathbf{M}_{tip}(\mathbf{r}) \chi_d^{-1}(\mathbf{r} - \mathbf{r}') \mathbf{M}(\mathbf{r}') \quad (S3)$$

$$U_\chi = -\frac{1}{2}\int d\mathbf{r} d\mathbf{r}' d\mathbf{r}_1 d\mathbf{r}_2 \mathbf{M}_{tip}(\mathbf{r}) \chi_d^{-1}(\mathbf{r} - \mathbf{r}_1) \chi(\mathbf{r}_1, \mathbf{r}_2) \chi_d^{-1}(\mathbf{r}_2 - \mathbf{r}') \mathbf{M}_{tip}(\mathbf{r}') . \quad (S4)$$

As MFM measures changes in the oscillation frequency of the tip, it is proportional to $-\partial_z^2 U$. Thus, we can define both the magnetization signal and susceptibility signal

$$I_m = -\partial_z^2 U_m, \quad I_\chi = -\partial_z^2 U_\chi. \quad (S5)$$

By subtracting and adding the MFM signal for a reversed tip, one can measure these two contributions separately, c.f. Supplementary Fig. 9, 10 and 11.

The relative strength of $I_m$ and $I_\chi$ depends on the magnetization of the MFM tip, $\mathbf{M}_{tip}(\mathbf{r})$. In our case the tip is of pyramidal shape and covered by a thin magnetic layer. Unfortunately, the domain structure of the magnetization is not known. We will therefore discuss below two models for the tip magnetization, called "dipolar tip" and "pyramidal tip".

For the dipolar tip, we simply approximate the tip by a point-like dipole at position $\mathbf{r}_0$ as $\mathbf{M}_{tip}(\mathbf{r}) = \mathbf{m}_{tip}\delta^3(\mathbf{r} - \mathbf{r}_0)$. For the pyramidal tip, in contrast, we assume that the sides of a pyramidal tip are covered with a thin magnetic layer of uniform width with a magnetization oriented in the $z$ direction. We assume that both the thickness of the magnetic layer and the rounding of the tip of the pyramid are much smaller than the distance $z_0$ of the tip from the surface and that the height of the pyramid is much larger than the lift height $z_0$. Both conditions are met in our experiment. This allows to ignore the rounding and to approximate the tip as an infinitely long pyramid. The precise value of the opening angle of the pyramidal tip and the number of sides of the pyramid are not important for the following qualitative discussion which focusses on the qualitative dependence of $I_m$ and $I_\chi$ on the distance $z_0$ of the tip from the surface. Technically, all integrals in supplementary equations (3) and (4) are evaluated using scaling arguments. In each case we checked numerically the validity of the scaling argument and the convergence of the integrals.

**Magnetic signal.** We consider a situation where the magnetization oscillates (approximately) periodically in one direction (we use the x-direction below) with period $\lambda$. In the helical phase, $\lambda$ is simply the wavelength of the helix but along a curvature domain wall we have a larger (projected) wavelength $\lambda_p = \lambda/\cos(\phi)$. We study the case $2\pi z_0 > \lambda$ which is of relevance for both the helical phase and curvature domain walls. For the following qualitative analysis, it is useful to consider the Fourier transform of $\chi_d^{-1}$ parallel to the surface of the sample, which decays exponentially with $z_0$

$$\chi_d^{-1}(q_{||}, z_0) = \int dr_{||} e^{-iq_{||}r_{||}} \chi_d^{-1} \sim e^{-|q_{||}|z_0}. \quad (S6)$$

In the helical phase and at curvature domain walls, the average magnetization vanishes. Therefore, the magnetization only has Fourier components which are multiples of $2\pi/\lambda$ (or $2\pi/\lambda_p$) and thus the oscillating magnetization signal is exponentially suppressed

$$I_m^{osc} \propto \cos\left(\frac{2\pi}{\lambda}x_0\right)e^{-\frac{2\pi}{\lambda}z_0} \quad (S7)$$

both for the pyramidal tip and the dipolar tip. At a curvature domain wall, $\lambda$ has to be replaced with $\lambda_p > \lambda$, resulting in a slower decay of the signal with the lift height $z_0$, see Supplementary Fig. 3.

Close to more complex textures, in contrast, there can be a net average magnetization along the domain wall and, hence, the $q_{||} = 0$ Fourier components contribute. In this case we obtain

$$I_m \propto \begin{cases} \dfrac{1}{z_0^3}, & \text{dipolar tip} \\ \dfrac{1}{z_0}, & \text{pyramidal tip} \end{cases}. \quad (S8)$$

The two-dimensional integral over the surfaces of the pyramid is responsible for the slower decay of its signal by two powers of $z_0$. For a rod-like defect oriented perpendicular to the surface (i.e., realizing a point-like defect on the surface), the signal decays with one power of $z_0$ higher

$$I_m \propto \begin{cases} \dfrac{1}{z_0^4}, & \text{dipolar tip} \\ \dfrac{1}{z_0^2}, & \text{pyramidal tip} \end{cases}. \quad (S9)$$

Here we assumed that $z_0$ is larger than the width of the defect.

**Susceptibility signal.** The susceptibility $\chi$ can also be split into a constant part $I_\chi$ and an oscillating part $I_\chi^{osc}$. In the helical phase without defects, by power-counting, we find that the non-oscillating part of the susceptibility signal yields a contribution

$$I_\chi \propto \begin{cases} \dfrac{1}{z_0^5}, & \text{dipolar tip} \\ \dfrac{1}{z_0}, & \text{pyramidal tip} \end{cases}. \quad (S10)$$

Note that the two signals differ by four powers of $z_0$ as one has to integrate in supplementary equation (4) twice over the two-dimensional surface of the MFM tip. The difference $\Delta I_\chi = I_\chi^{dw} - I_\chi^h$ of the signal on top of the domain wall, $I_\chi^{dw}$, and in the helical phase, $I_\chi^h$, decays with one power of $z_0$ faster (assuming again that $z_0$ is larger than the width of the domain wall)

$$\Delta I_\chi \propto \begin{cases} \dfrac{1}{z_0^6}, & \text{dipolar tip} \\ \dfrac{1}{z_0^2}, & \text{pyramidal tip} \end{cases}. \quad (S11)$$

By analyzing supplementary equation (8), one can show that the oscillating part of $I_\chi$ which arises from the oscillating part of the susceptibility decays exponentially

$$I_\chi^{osc} \propto \cos\left(\frac{2\pi}{\lambda} x\right) p(z_0) e^{-\frac{2\pi}{\lambda} z_0}, \quad (S12)$$

similar to $I_m^{osc}$ but with a much smaller ($z_0$-dependent) prefactor $p(z_0)$.

**Conclusions.** Our analytical analysis qualitatively explains our main experimental and numerical observations. The experiment operates in a regime where the exponentially suppressed oscillatory magnetization signal $I_m^{osc}$ is of similar size compared to the susceptibility signal, $\Delta I_\chi$. For larger distances $z_0$ between the tip and the sample, the susceptibility signal dominates close to curvature domain walls which do not have an average magnetization. The exponentially suppressed oscillating contrast from the susceptibility signal, $I_\chi^{osc}$, is never observable in our experiments and barely visible in the numerics (see Supplementary Fig. 9). Non-periodic features or features with a period much larger than $z_0$ can, however, easily be resolved, see, e.g., Supplementary Fig. 9.

In Supplementary Fig. 1 the $z_0$ dependence of the MFM signal is shown in three cases. First, on top of the helical background, we observe an exponential decay of the oscillatory part of the signal, see supplementary equation (7). The fitted decay length is nominally about 50% larger than predicted, probably due to the limited fitting range and the relatively small values of $z_0$ available for the fit. In regions 2 and 3 of Supplementary Fig. 1 a much slower decay of the signal is observed. In region 2, where the magnetization can locally be described by a curvature domain wall, the signal arises most likely from the susceptibility. The fitted decay $\propto \frac{1}{z_0^\alpha}$ with $\alpha \approx 2.5 \pm 0.5$ appears to be slightly faster than the expected $\frac{1}{z_0^2}$ for an idealized pyramidal tip, supplementary equation (11). The discrepancy can either arise due to the rather broad and more complex magnetic structure of the sample or from possible magnetic domains in our tip magnetization. Finally, in region 3, we do not expect a pure power law due to the rather complex magnetic structure which, most likely, is also characterized by a net magnetization. Therefore both $I_\chi$ and $I_m$ may contribute to the signal. The fitted power law with exponent $\alpha \approx 1.66 \pm 0.33$ consistent with $\frac{1}{z_0^2}$ predicted both for the magnetization signal of a rod-like defect, supplementary equation (9), for the susceptibility contrast of a domain wall, supplementary equation (11).